\documentclass[conference]{IEEEtran}
\pdfoutput=1
\usepackage{cite}
\usepackage{amsmath,amssymb,amsfonts}
\usepackage{algorithmic}
\usepackage{graphicx}
\usepackage{textcomp}
\usepackage{xcolor}
\usepackage{multirow}
\def\BibTeX{{\rm B\kern-.05em{\sc i\kern-.025em b}\kern-.08em
    T\kern-.1667em\lower.7ex\hbox{E}\kern-.125emX}}

\usepackage{url}
\usepackage{flushend}
\usepackage{hyperref}
\usepackage[caption=false,font=footnotesize]{subfig}
\usepackage{numprint}
\npthousandsep{,}

\usepackage{enumitem}
\setlist[itemize]{leftmargin=*}
\setlist[enumerate]{leftmargin=*}
\setlist{nolistsep}

\IEEEoverridecommandlockouts
\IEEEpubid{\makebox[\columnwidth]{~\copyright2019 IEEE \hfill} \hspace{\columnsep}\makebox[\columnwidth]{ }}

\begin{document}

\title{Parallelizing Training of Deep Generative Models on Massive Scientific Datasets} 

\author{Sam Ade Jacobs \quad Brian Van Essen \quad David Hysom \quad
    Jae-Seung Yeom  \quad Tim Moon  \quad 
    Rushil Anirudh\\
    Jayaraman J.\ Thiagaranjan  \quad Shusen Liu  \quad Peer-Timo
    Bremer  \quad  Jim Gaffney \quad Tom Benson \quad Peter
    Robinson\\\hfill Luc Peterson \quad Brian Spears\hfill\\
\vspace{2mm}
\IEEEauthorblockA{Lawrence Livermore National Laboratory}}

\maketitle
\IEEEpubidadjcol
\begin{abstract}
Training deep neural networks on large scientific data is a
challenging task that requires enormous compute power, especially if no
pre-trained models exist to initialize the process. We present
a novel tournament method to train traditional as well as generative
adversarial networks built on LBANN, a scalable deep learning framework
optimized for HPC systems.  LBANN 
combines multiple levels of parallelism and exploits some of the worlds
largest supercomputers.

We demonstrate our framework by creating a complex predictive model
based on multi-variate data from high-energy-density physics
containing hundreds of millions of images and hundreds of millions
of scalar values derived from tens of millions of simulations of inertial
confinement fusion. Our approach combines an HPC workflow and extends LBANN
with optimized data ingestion and the new tournament-style training
algorithm to produce a scalable neural network architecture using a
CORAL-class supercomputer. Experimental results show that 64 trainers (1024 GPUs)
achieve a speedup of $70.2\times$ over a single trainer (16 GPUs) baseline, 
and an effective $109\%$ parallel efficiency. 
\end{abstract}


\newenvironment{item_tight}{
\begin{itemize}
  \setlength{\itemsep}{1pt}
  \setlength{\parskip}{0pt}
  \setlength{\parsep}{0pt}
}{\end{itemize}}


\section{Introduction}

The explosion of deep learning in recent years has unlocked the potential to
change the way in which we tackle large scale scientific simulations.  While
machine learning techniques are being applied to many academic, medical, and
commercial applications, the field of large scale, scientific machine learning
techniques is just beginning to emerge. The scientific community is leveraging
advances in deep learning and computational workflows to bridge the gap between
approximate scientific simulations and more accurate but expensive and
often limited
experimental analysis.  We refer to this coupling of traditional scientific
computing with deep learning as cognitive simulation 
and envision it as a new way to enable
predictive science.  There are many challenges to this new methodology 
including 
generating large scale scientific data sets for training,
developing learning methods for unlabeled data, scaling up deep learning training methods
to leverage state of the art supercomputers, and developing novel neural network
architectures for scientific applications.  In this paper we present our
approaches 
that address these challenges, with a particular focus on the
development of parallel methods for large scale training of generative models.

The overarching goal of our cognitive simulation research is 
the development of
techniques that can use machine learning to augment workflows and to supplement, combine, or
replace existing heuristics.  Depending on where these models engage the
simulation they are referred to as being 1) in-the-loop, 2) on-the-loop, or 3)
around-the-loop~\cite{spears-deeplearning}.  Notionally, examples of these levels of
engagement are 1) directly in a physics simulation, 2) observing and influencing
a simulation code, or 3) part of the simulation campaign.  Using machine learning (or specifically deep learning) at each
of these levels has different requirements for the size of admissible models and
the required speed of inference.  In this paper we present our work on
developing novel generative models that take physical constraints into
account and are used 
around-the-loop of an inertial confinement fusion (ICF)
simulation campaign. Here we use results from a recently proposed semi-analytic simulation model~\cite{gaffney2014thermodynamic, springer2013integrated}.
to simulate the behavior of the implosion of a fuel capsule in instruments such as the
National Ignition Facility (NIF) at Lawrence Livermore National Lab (LLNL).
These types of simulations are critical to understanding the physics occurring
during an implosion and  
subsequently
improving the output of the experimental campaigns.

Scientific machine learning presents the researcher with several unique
challenges.
First and foremost, scientific data sets that come from
large-scale simulations or experiments can be extremely large, are frequently
high dimensional, and require domain expertise to label the data. Secondly,
because we 
are applying novel analysis to uncommon data
there is a lack of predefined neural network models or proven model families to draw
upon when starting a new analysis.  As such, new neural network architectures
have to be designed and tuned for specific problems and data sets. This requires
massive training of hyperparameters and model exploration.
 
In this paper, we present our work on extending a scalable deep learning
framework with a novel tournament parallel algorithm that is able to train
large, complex generative models on massive amounts of multimodal data.
Additionally, we have developed a novel, distributed, in-memory data store that is optimized
for minimizing file system access during training of deep neural networks.  We
integrated these algorithms and capabilities into the Livermore Big Artificial
Neural Network (LBANN) deep learning framework. LBANN is an open source
HPC-centric framework built on the Hydrogen distributed linear algebra 
and Aluminum GPU-aware communication libraries \cite{van2015lbann, lbann2019,
  hydrogen2019, dryden2018, aluminum2019}.  The novel tournament algorithm,
Let a Thousand Flowers Bloom (LTFB), primarily targets scaling up the training
of deep neural networks on massive data sets and leveraging
leadership-class HPC systems. The main thrust of LTFB is to minimize the amount of
synchronization required for each step of the training algorithm and to develop
a mechanism for combining independently trained models. 
The key feature of LTFB is the
independent training of models on partitioned, more manageable
data sets, while yielding a model that is as good a one trained on the entire
data set. This allows for strong scaling and is accomplished by periodically
running a tournament where a locally trained model competes with other models on a
held-out ``tournament'' data set. The winning model continues and losing models are
discarded.  Propagation of the winning model serves as an efficient encoding of
key features from other parititons of the data.  The LTFB algorithm provides a mechanism for scalable data ingestion
by allowing partitioning of the data set without loss of generalizability.
Furthermore, it enables scalable exploration of the initial state space as well
as the state space after each tournament.  Implementing LTFB within the LBANN
framework and coupling it with the data-store allowed us to efficiently train on a 2TB database of
10 million 5-D input parameters, 120 million multispectral images and 10 million
15-D scalar values. 
This is the first demonstration of learning at such scale on a multi-variate scientific data set.

In addition to our work on large scale learning, developing a cognitive
simulation capability requires innovation in scientific workflows and neural
network architectures.  We briefly present an overview of these capabilities in
Sections \ref{sec:scienceml}.  
The LTFB algorithm and data-store
are discussed in Section \ref{sec:lbann}, followed by
our experiments in Section \ref{sec:exp}, and related work in Section
\ref{sec:related}.  In summary, we show that LTFB provides the ability to strongly
scale the training time of a single neural network architecture, taking
advantage of thousands of GPUs, while maintaining a reasonable sized mini-batch.  This
capability provides a unique ability to explore complex neural network
architectures, while using unsupervised learning methods on massive data sets.
We summarize our contributions in this paper as follows:
\begin{itemize}
\item A scalable deep learning framework that is able to strongly scale training
  of a single model on thousands of GPUs, while only requiring modest data
  parallelism.
\item A novel tournament method, LTFB, that is optimized for complex generative
  models, minimizes communication, and enables efficient partitioning of large
  data sets while maintaining model generalizability.
\item A new in-memory distributed data-store optimized for training deep neural
  networks, which leverages low-latency, high-bandwidth interconnect for
  efficient file system access.
\item First demonstration of scalable training of complex cyclic generative models to
1024 GPUs without loss in quality. 
\end{itemize}

\section{Scientific Problem}
\label{sec:scienceml}
\subsection{Objective}
\begin{figure}[]
  \centering
  \includegraphics[width=0.5\textwidth]{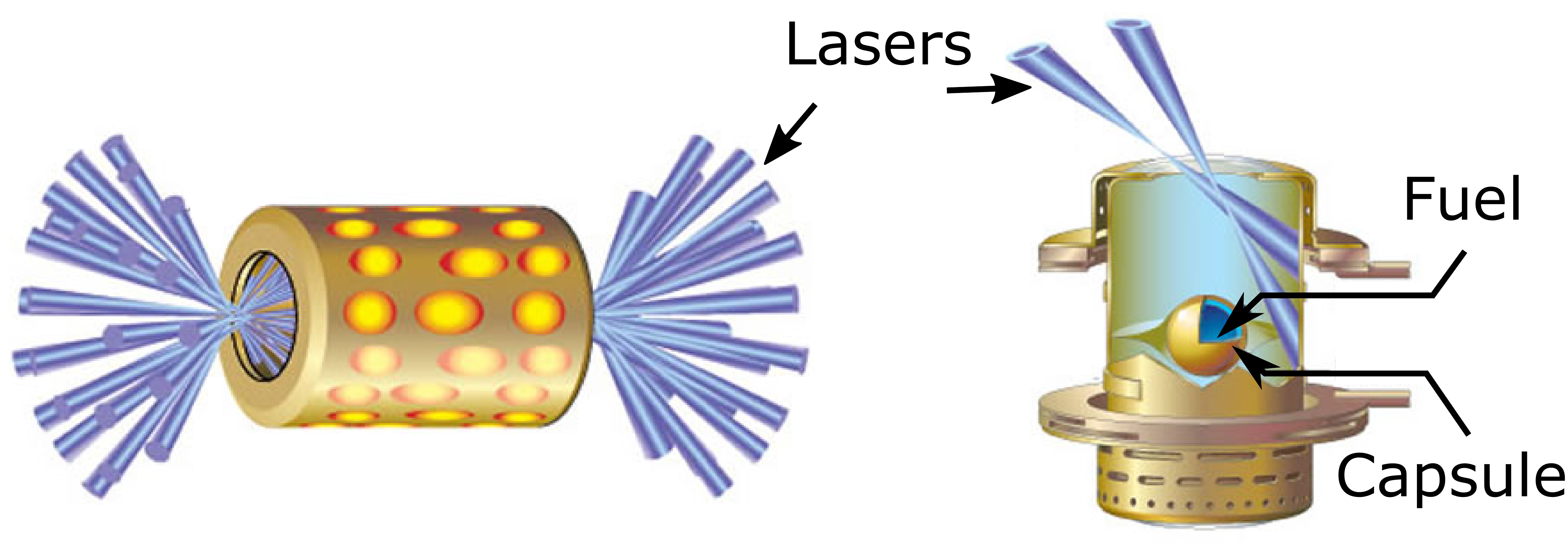}
  \caption{Schematic of an ICF experiment at NIF. High energy lasers
    heat and compress a target capsule containing thermonuclear fuel,
    resulting in a fusion process.}
  \label{fig:ICF_capsule}
\end{figure}

This work was developed to support inertial confinement fusion (ICF)
experiments at the National Ignition Facility (NIF). These experiments
use the world's largest laser to heat and compress a millimeter-scale
target filled with frozen thermonuclear fusion fuel (see
Figure~\ref{fig:ICF_capsule}). Under sufficiently well-controlled
conditions, the compressed fusion fuel will produce enough energy for
the target to self-heat, leading to a runaway implosion process called
ignition. Ultimately, ignition will produce more fusion energy than
the driving laser energy \cite{betti2015alpha}, enabling study of
fusion energy production, astrophysical phenomena, and nuclear weapons
processes.

Our particular goal was to develop a fast surrogate model that can
predict the outcome of ICF experiments. It could be used, for
instance, for experiment optimization, statistical uncertainty
quantification, or efficient sampling of the experimental parameter
space. Robust model inversion could also be used to infer the physics
processes underlying experimental observations. Each ICF experiment
generates a rich signature set composed of scalars and images, so the
surrogate model would be a large generative model for complicated
multimodal data. Due to the complexity and sensitivity of ICF
experiments, a large volume of simulated data was necessary to train
such a model.

\subsection{JAG simulator for ICF}

First-principles simulation of ICF implosions require high-fidelity
multiphysics simulations \cite{marinak01}, frequently taking thousands
of CPU-hours per sample. To produce the required volume of data, we
instead used the JAG model, a semi-analytical model for the final
stages of an ICF implosion \cite{gaffney2014thermodynamic}. Since all
experimental signals are generated during this final period, JAG is
capable of generating a realistic set of multi-modal outputs while
only taking a few CPU-seconds. We performed simulations over a
5-dimensional parameter space --- controlling the strength of the
laser drive and the 3D shape of the imploding shell --- and simulated
X-ray cameras on three different lines of sight --- each with
4-channel hyperspectral energy resolution and spatial resolution of
\(64\times 64\) pixels. We also postprocessed the JAG output to obtain
15 scalar-valued observable signatures. Thus, each data sample is a
pair consisting of an input 5-vector and an output bundle of 15
scalars and 12 images. Generally, varying the drive parameters
resulted in highly non-linear variations in the scalar performance
metrics and varying the shape parameters resulted in major changes in
the X-ray images.

\subsection{Ensemble workflow}

Given the strong non-linearities in ICF experiments, it was important
to densely cover the five-dimensional parameter space for JAG
simulations. This was challenging not only because of the large number
of runs and corresponding files, but also because JAG is so
fast. Running JAG and performing postprocessing only takes about a
minute, so a workflow system's runtime can be dominated by the
overhead of scheduling, placing, and executing jobs. We addressed this
problem with an extension of the Merlin workflow system
\cite{doi:10.1063/1.4977912}, which uses a custom combination of
various open-source components to build a highly flexible and
efficient framework. We used a spectral sampling approach to optimally
assign simulation parameters \cite{kailkhura2018spectral}, resulting
in 10 million simulations for the training dataset and 1 million for
the test dataset. To manage this enormous number of samples, we
packaged the data into \numprint{10000} HDF5 files, each of which contains \numprint{1000}
samples.

\subsection{Neural network architecture}

\begin{figure}[]
\centering
  \includegraphics[width=\columnwidth]{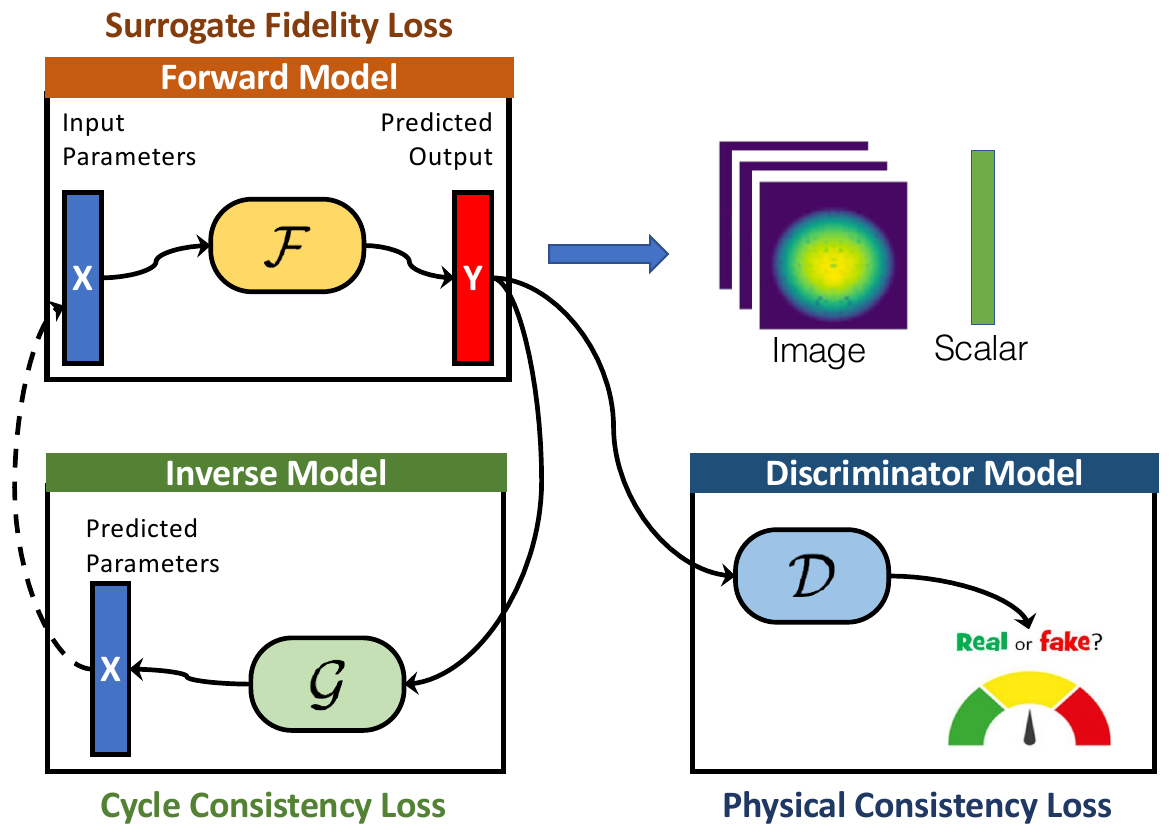}
  \caption{CycleGAN surrogate model for ICF experiments.}
\label{fig:cycleGAN}
\end{figure}

The surrogate model was implemented as the CycleGAN shown in Figure
\ref{fig:cycleGAN} since it imposes several desirable consistency
conditions. First, \textit{internal consistency} means that the
forward model predicts all of the output modalities jointly. This
avoids the uncorrelated errors and physically invalid solutions that
can arise if each output modality is predicted independently. Second,
\textit{physical consistency} means that predictions are realistic,
ideally to the point where they are statistically identical to the
training data. We approximate this by training an adversarial
discriminator model to distinguish predictions from data samples
\cite{Goodfellow2014}. This is technically not a physical constraint,
but it does markedly improve the quality of predicted images. Third,
\textit{self consistency} means that there is an inverse model that
projects back to the original input. This is particularly useful since
both the forward and inverse models would be useful for domain
scientists in ICF. From a machine learning perspective, consistency
between the forward and inverse models also acts as regularization on
an otherwise highly underdetermined problem.

More precisely, the forward model
\(\mathcal{F}:\mathbb{R}^5\rightarrow\mathbb{R}^{20}\) maps from the
5-D experiment parameter space to a 20-D latent space. This is trained
a priori using a multimodal autoencoder of all outputs. Predicted
scalar values and X-ray images are obtained by passing latent space
vectors into a decoder network and the internal consistency condition
is enforced with the mean absolute error loss. The discriminator model
\(\mathcal{D}:\mathbb{R}^{20}\rightarrow \{0, 1\}\) is trained
adversarially. The inverse model
\(\mathcal{G}:\mathbb{R}^{20}\rightarrow\mathbb{R}^5\) attempts to
enforce the self consistency condition
\(\mathcal{G}\circ\mathcal{F}\approx\mathcal{I}\) using the mean
absolute error loss. Each of these components is implemented as a
standard fully-connected neural network. A complete description of the
network is available at \cite{osti_1510714}.

\section{Scalable Deep Learning}
\label{sec:lbann}

We used LBANN \cite{van2015lbann, lbann2019}, an open-source deep learning framework
from LLNL, as the platform for this research. LBANN implements a suite
of algorithms for training deep learning models on distributed memory
architectures, exploiting both data- and model-parallelism to achieve
scalable performance on HPC systems. We have extended LBANN in this
work to support GANs and have also upgraded the internal
infrastructure to improve robustness and scalability when training
multiple concurrent models. We have developed a distributed in-memory
data store that minimizes access to the parallel file system during
training. Finally, we have extended the LTFB algorithm presented by
Jacobs et al. \cite{Jacobs2017} to handle GANs.

\begin{figure}[]
  \centering
  \includegraphics[width=0.5\textwidth]{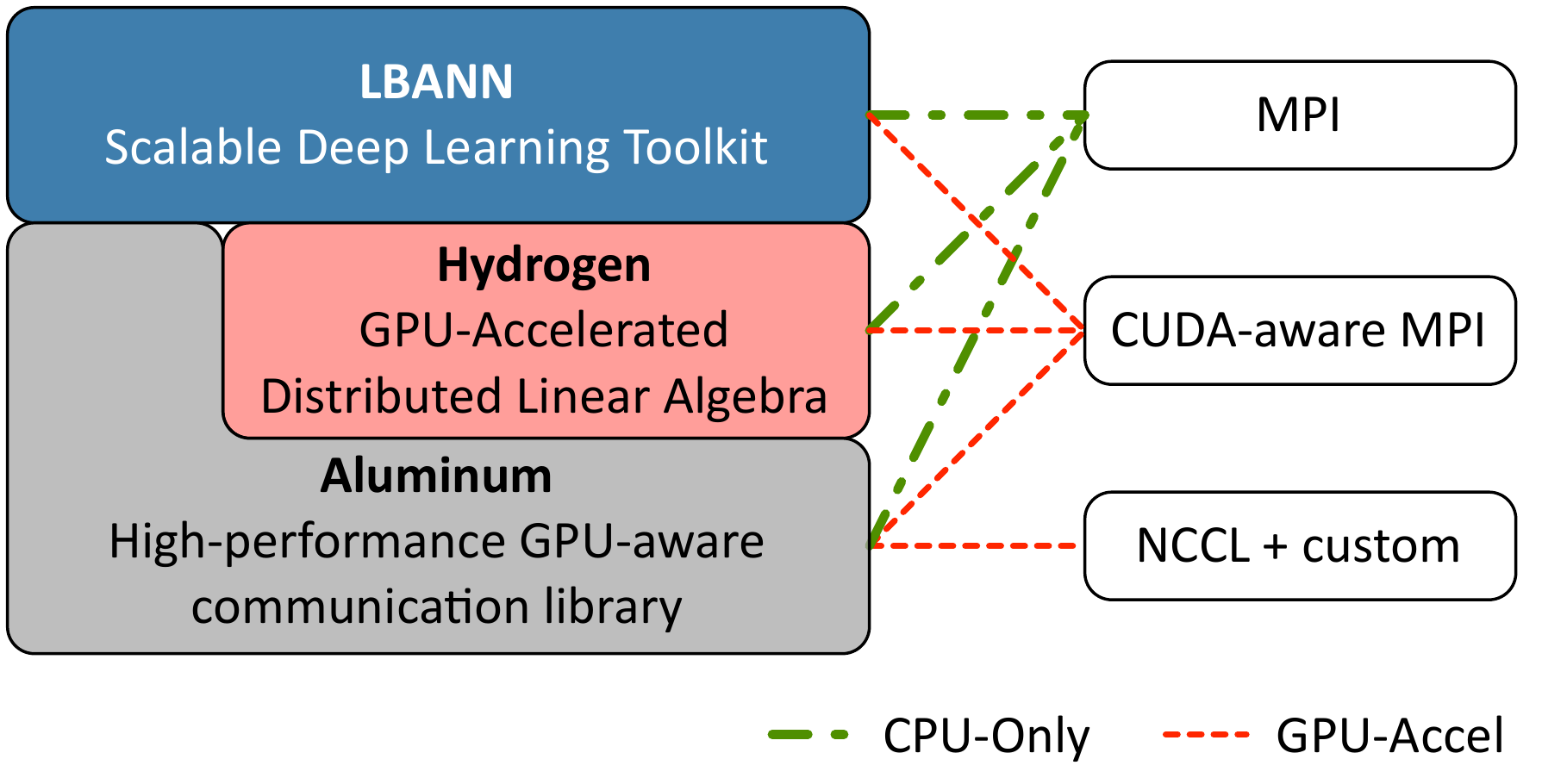}
  \caption{ LBANN's software stack, consisting of the Hydrogen linear
    algebra library and the Aluminum communication library. }
  \label{fig:lbann_software_stack}
\end{figure}

LBANN's software stack is shown in
Figure~\ref{fig:lbann_software_stack}. The top-level framework is
written in C++ and CUDA. Hydrogen \cite{hydrogen2019}, a fork of
Elemental \cite{poulson2013elemental}, provides distributed linear
algebra with GPU acceleration and Aluminum provides GPU-aware
asynchronous communication \cite{aluminum2019}.

\subsection{Trainers and models}
\label{sec:lbann:lbann}

Two key concepts in LBANN are those of \textit{trainers} and
\textit{models}. A trainer is a collection of compute resources that
operate together as a unit. A model is a neural network, comprised of
a directed acyclic graph (DAG) of tensor operations (``layers''),
trainable parameter tensors (``weights''), and data
readers. Naturally, trainers are responsible for training models,
usually with a variant of stochastic gradient descent. During a
model's mini-batch step, the trainer will ingest and preprocess data
with a data reader and pass it into the model's DAG. Each trainer
manages one or more models and it may accelerate computation with
data-parallelism, model-parallelism, or both. Observe that running
LBANN with multiple trainers results in two levels of parallelism:
within each trainer and between trainers.

\begin{figure}[]
  \centering
  \includegraphics[width=0.5\textwidth]{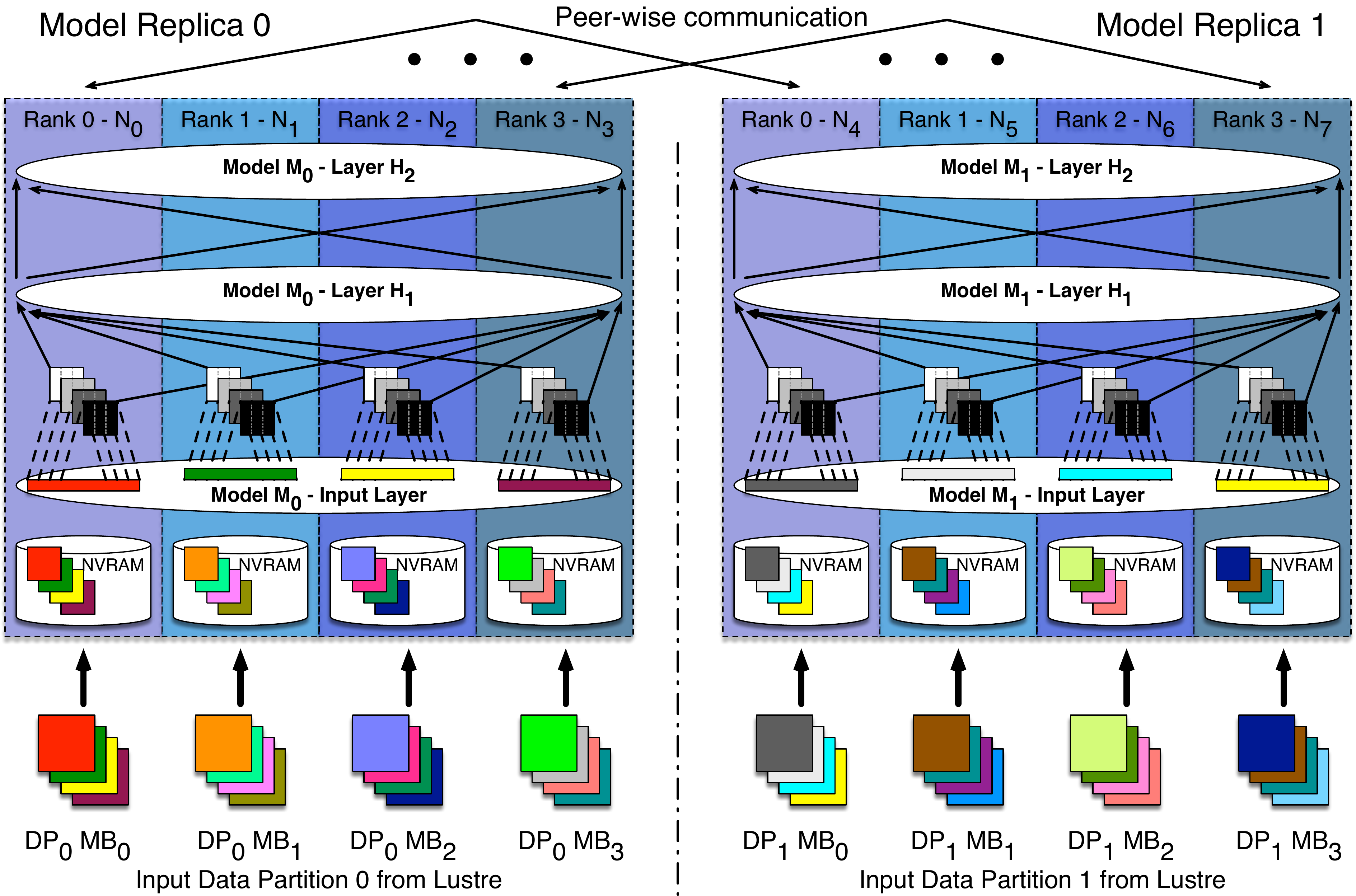}
  \caption{Example of LBANN running with two trainers, each of which
    consists of four MPI ranks.}
  \label{fig:lbann_parallelism}
\end{figure}

\subsection{Data store}
\label{sec:lbann:datastore}

One of the key challenges of this work was training data
management. Typically, data from large simulation runs are generated,
triaged, and dumped to a parallel file system like Lustre or GPFS. A
naive data reader would then populate each mini-batch by opening the
files for its required data samples. However, recall from
Section~\ref{sec:scienceml} that each data file consists of \numprint{1000}
samples. In addition, samples should be drawn randomly from the
dataset to make sure they are representative of the source
distribution. Consequently, each process will open many files and each
file may be accessed by multiple processes at the same time. This puts
a huge burden on the file system and drastically slows down data
ingestion.  In a naive approach to I/O, data ingestion can dominate the training
time of the model.

To avoid this problem, we have implemented a scalable, distributed in-memory data
store in LBANN. Each process in a trainer is assigned to manage a subset of the
data to be cached in system memory.  Therefore the total data store capacity is
proportional to the number of compute nodes in the trainer. At each training
step, processes will distribute the locally-cached samples that they manage to
the trainer's MPI rank
that requires them for the upcoming mini-batch.
This shuffling is done with non-blocking communication
on background threads, so it efficiently overlaps with other
computation. The data store itself utilizes Conduit to provide a
data-type-agnostic in-memory framework for managing data samples
\cite{conduit2019}. While some popular open-source frameworks provide
local data caching on a single node \cite{tensorflowDataCache, pytorchDataLoader}, this work presents --- to the
best of our knowledge --- the first distributed in-memory data store
in a deep learning framework.

We have explored two approaches to populate this data store, dynamic and preloading. In the
dynamic approach, data samples are read from data files during the
first training epoch in a similar manner as naive data ingestion, but
samples are cached in the data store as they are used. Thus, we only expect to
suffer the previously discussed performance penalties during the first
epoch, after which no data is read from the file
system.  For each subsquent epoch, data is incrementally shuffled between the ranks at each
mini-batch step. 
 Preloading is a second optimization technique that fully populates the data
store prior to training.  While this shifts the file I/O burden to a
preprocessing step, it allows for optimal access to samples from multi-sample
file formats such as HDF5.  To preload the data store, each process is assigned a disjoint subset of
the data files and accesses, in parallel, all of the data samples within each file. This
minimizes the number of files each process opens concurrently, and ensures that each
file is only opened by one process per trainer. During training itself, no data is
read from the file system.

\begin{figure}[]
  \centering
  \includegraphics[width=0.5\textwidth]{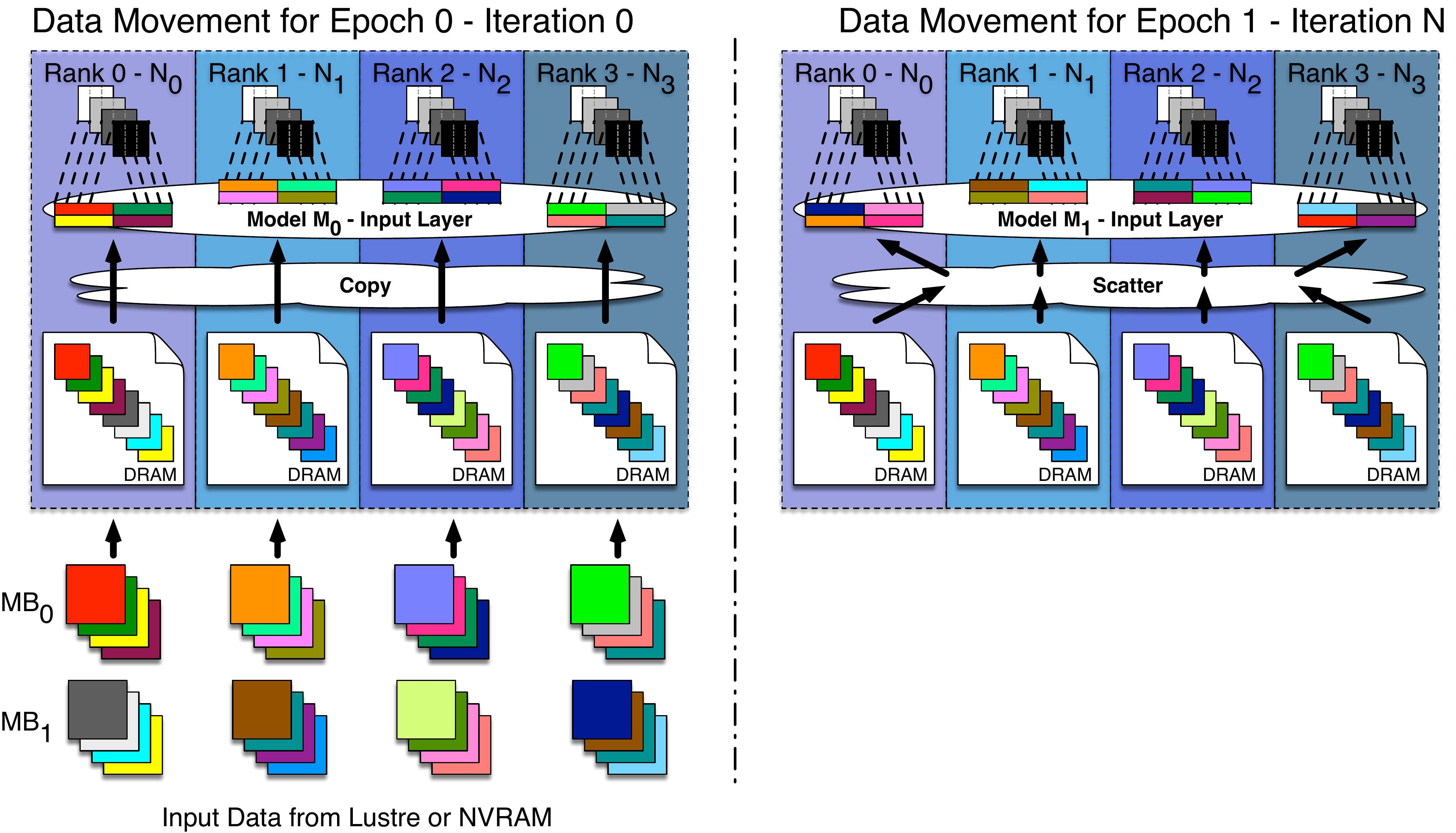}
  \caption{Example of an LBANN data store for a trainer with four MPI
    ranks. The data store is populated dynamically during the first
    training epoch, after which no data is read from the file system.}
  \label{fig:lbann_data_store}
\end{figure}

\begin{figure*}[]
  \centering
  \begin{minipage}{0.75\textwidth}\centering
    \subfloat[Trainers independently train GANs using a partition of
    the training dataset.]
    {\includegraphics[width=0.95\textwidth]{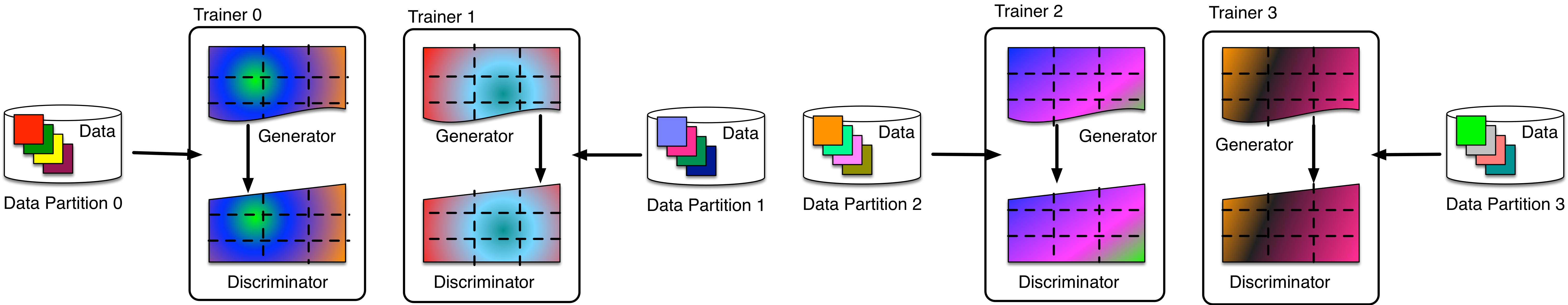}}
  \end{minipage} \medskip \\
  \begin{minipage}{0.75\textwidth}\centering
    \subfloat[During an LTFB round, trainers will pair up, exchange
    their generators, and evaluate them against their local
    discriminators. The better generator is retained for further
    training.]
    {\includegraphics[width=0.95\textwidth]{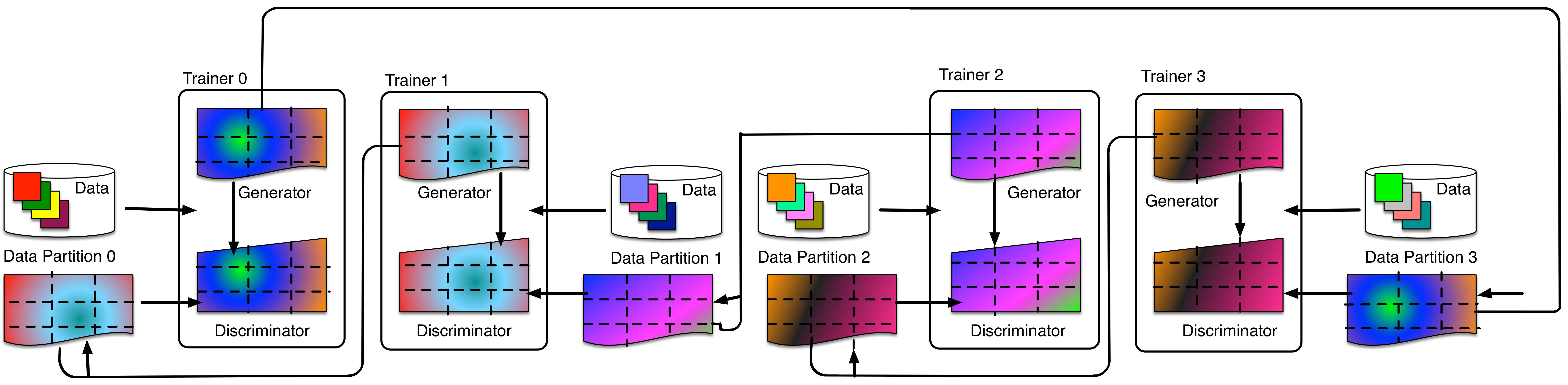}}
  \end{minipage}
  \caption{ Example of training GANs with LTFB using four trainers. Dashed lines
  in each trainer represent that the models (or data) are internally parallelized across GPUs.}
  \label{fig:ltfb_gan}
\end{figure*}

\subsection{LTFB for GANs}
\label{sec:lbann:ltfb}

``Let a Thousand Flowers Bloom'' (LTFB) is a decentralized variant of
population-based training \cite{Jacobs2017}
\cite{DBLP:journals/corr/abs-1711-09846}, primarily intended to
achieve scalable performance when training neural networks with
massive datasets on HPC systems.  It begins by initializing multiple
trainers and partitioning the training dataset between them. The
trainers construct models and train them in a loosely-coupled manner.  Training
each model between tournaments is independent, and the tournament provides
periodic coupling between the trainers.  The models are
initialized with different weights and hyperparameters, but even
identical models will diverge over the course of training since
trainers expose different silos of data.  Periodically, e.g.\ at
predefined mini-batch intervals, trainers are randomly paired up and
made to exchange models. Each trainer will evaluate its two models on
a local tournament data set, keeps the one that achieves a better evaluation
metric, and then resumes training. Even though each trainer only exposes a
model to a subset of the data, models that survive LTFB are likely have been exposed
to many trainers at different times, and thus are expected to capture
the characteristics of the entire dataset.

This approach has some nice scaling properties. Communication between
trainers is limited to infrequent peer-to-peer model exchanges, so
increasing the number of trainers does not incur significant
performance overheads. Furthermore, if a trainer is efficiently mapped to the
hardware topology, e.g.\ to a compute node, intra-trainer
communication is well optimized and the number of processes is typically
small enough to fruitfully apply strong-scaling techniques. Finally,
the fact that the dataset is divided amongst the trainers helps keep
file I/O scalable.

Previous work on LTFB demonstrated modest scaling on image
classification benchmarks \cite{Jacobs2017}. We extended that work to
GANs by only exchanging generator models during LTFB rounds and
keeping discriminators local to each trainer, as shown in
Figure~\ref{fig:ltfb_gan}. This approach intuitively mimicks the
practice of educating a student with multiple teachers, which has been shown
to improve the quality of trained generative models on the MNIST,
CIFAR-10 and CelebA
benchmarks~\cite{DBLP:journals/corr/DurugkarGM16}. It also reduces the
inter-trainer communication volume during each LTFB round.

\section{Experiments}
\label{sec:exp}
\begin{figure}[ht]
  \centering
  {\includegraphics[width=0.48\textwidth]{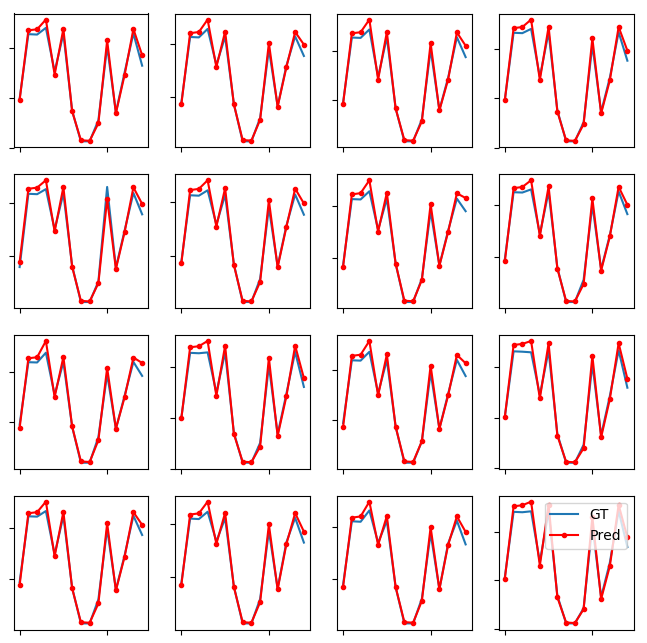}}
  \caption{Ground truth and LTFB CycleGAN predicted 15-D scalars for 16 validation
samples.  Note that the ground truth in blue is mostly covered by the GAN's
prediction in red.}
  \label{fig:ltfb_pred_scalar}
\end{figure}

\begin{figure*}[ht]
  \centering
  \begin{minipage}{0.16\textwidth}\centering
    \subfloat[Ground truth at view0 channel0\label{fig:gt_v0c0}]
    {\includegraphics[width=0.95\textwidth]{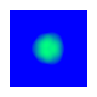}}
  \end{minipage}
  \begin{minipage}{0.16\textwidth}\centering
    \subfloat[Predicted at view0 channel0\label{fig:pred_v0c0}]
    {\includegraphics[width=0.95\textwidth]{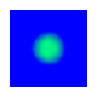}}
  \end{minipage}
  \begin{minipage}{0.16\textwidth}\centering
    \subfloat[Ground truth at view1 channel1\label{fig:gt_v1c1}]
    {\includegraphics[width=0.95\textwidth]{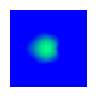}}
  \end{minipage}
  \begin{minipage}{0.16\textwidth}\centering
    \subfloat[Predicted at view1 channel1\label{fig:pred_v1c1}]
    {\includegraphics[width=0.95\textwidth]{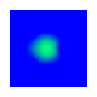}}
  \end{minipage}
  \begin{minipage}{0.16\textwidth}\centering
    \subfloat[Ground truth at view2 channel2\label{fig:gt_v2c2}]
    {\includegraphics[width=0.95\textwidth]{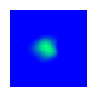}}
  \end{minipage}
  \begin{minipage}{0.16\textwidth}\centering
    \subfloat[Predicted at view2 channel2\label{fig:pred_v2c2}]
    {\includegraphics[width=0.95\textwidth]{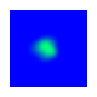}}
  \end{minipage}
  \caption{Capsule images (at selected views and channels) from the JAG model's
    output (ground truth) and generated by the LTFB CycleGAN generator network.}
  \label{fig:gan_network_images}
\end{figure*}

The LTFB algorithm enables a new dimension for strong scaling deep learning
training.  We achieve strong scaling by allowing the global data set to be split
into smaller partitions and still yielding a trained model that generalizes well
on the entire testing data set.  The critical contribution of the LTFB training
method is that the winning model has learned enough from each trainer and
associated data partition to accurately capture the characteristics of the
entire training data set.  Essentially, when a model is shared it is acting as
an encoded representation of its data partition.  Thus transfering that learned
representation to the new trainer for continuing education.  As we will demonstrate in these experiments, the
ability for LTFB to strong scale with a given mini-batch size allows it to
unlock a new axis of parallelism that is independent of, and composable with
existing data and model parallel methods.

 To demonstrate the efficacy of the LTFB algorithm we run multiple experiments
that show the limits of strong scaling via data parallelism, the challenges of
balancing data ingestion and parallel training, a comparison of a plain ensemble
method, and the ability of LTFB to provide strong scaling while producing a
model of equivalent or better performance.  It is important to note that for
these experiments we worked with a consistent set of model architecture and
hyperparameters for the CycleGAN network described in Section
\ref{sec:scienceml}.  Over the course of developing the network and curating
the data set, 
we identified that a mini-batch size of $128$ samples, an \texttt{adam} optimizer,
and an initial learning rate of $0.001$ work well for these experiments.
Figures~\ref{fig:ltfb_pred_scalar} and \ref{fig:gan_network_images}
show some ground truth scalars and images
respectively alongside predictions from our model. 

\subsection{Setup}
\label{sec:exp:setup}
Our experiments are run on the Lassen supercomputer at Lawrence Livermore
National Laboratory, within the Livermore Computing
collaboration zone (LC CZ).  Lassen is a CORAL-class
system (like Sierra), with 795 nodes, each of which consists of two IBM POWER9 CPUs and
four Nvidia Volta V100 GPUs.  The GPUs and CPUs are interconnected with three
NVLINK2 connections and each Volta has 16GB of memory the node has 256GB of
system memory.  Nodes are interconnected via dual-rail InfiniBand EDR.  Our
implementation of LTFB and the data store were build on a recent development
version of LBANN, Hydrogen, and Aluminum.  Our software development environment
used GCC 7.3.1, Spectrum MPI 2019.01.30, CUDA 9.2.148, cuDNN 7.5.0, and NCCL
2.4.2. For all results we use single-precision floating point data types.

\subsection{Small-scale data parallelism}

\newcommand{\dataparallelincrease}{9.36}
\newcommand{\dataparallelefficiency}{58}

\begin{figure}
  \centering
  \includegraphics[width=0.5\textwidth]{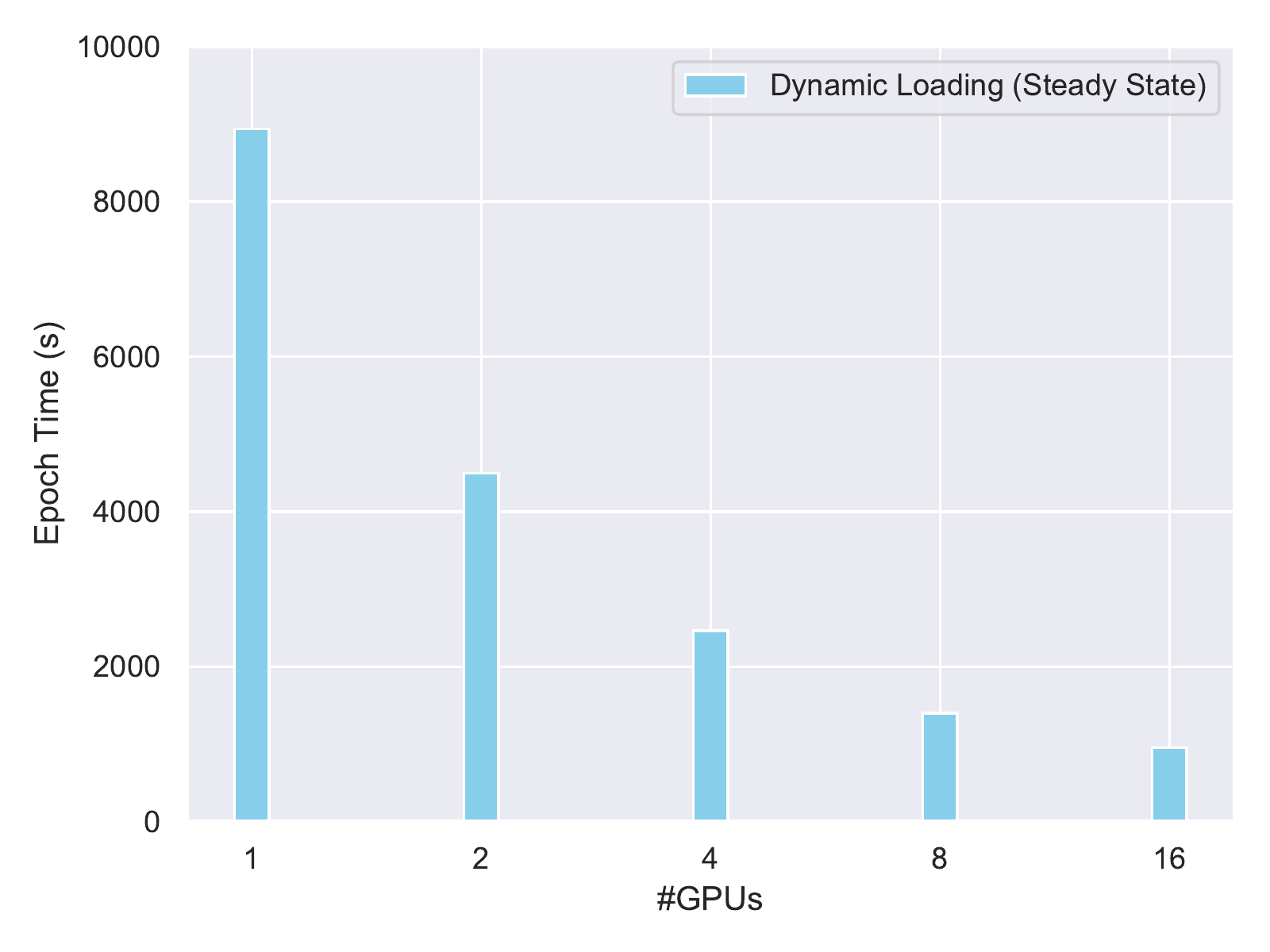}
  \caption{Evaluating the benefit of increasing the number of GPUs used for data
    parallel training a single model.}
  \label{fig:data_parallel_lassen}
\end{figure}

The most common method of strong scaling model training time is data
parallelism: where the samples within a mini-batch are distributed across
multiple nodes and partial error signals and gradient updates are aggregated via
all-reduce collectives during back propagation. The ability of this method to
scale is limited by the global mini-batch size and the amount of compute
required per node as it will become communication bound.  Figure
\ref{fig:data_parallel_lassen} shows the results of training the CycleGAN on
using only $1M$ samples from the full $10M$ sample data set.  With simple data
parallel scaling from 1 node and only 1 GPU training the CycleGAN to using 4 nodes
and 16 GPUs there is a ${\dataparallelincrease}\times$ improvement in steady state
epoch time.  Figure \ref{fig:data_parallel_lassen} does show that the benefits
of data parallel scaling are starting to diminish around 4 nodes and 16 GPUs,
with a decrease in parallel efficiency down to ${\dataparallelefficiency}\%$.
To provide a good balance between data parallel and independent trainers for the
tournament experiments, we use 4 nodes and 16 GPUs to train each model in
subsequent sections.

\subsection{Exploring the Data-store}

One of the challenges of working with large scale scientific data sets is that
the cost to ingest the training, evaluation, and testing data can easily become
a bottleneck or complete impediment to training the network.  As noted in
Section \ref{sec:scienceml}, the ICF data set is stored in \numprint{10000} HDF5 files with
\numprint{1,000} samples per file.  When the data was generated the samples were stored
in these HDF5 bundles in the order in which the 5-D input space was explored by
the semi-analytical simulation tool.  Therefore, to effectively train the neural
network it is necessary to randomly sample from all of the available files to
pull a uniform distribution for each mini-batch and thus each step of the Stochastic Gradient Descent (SGD)
algorithm.  This type of access pattern creates a significant burden on the
parallel file system and leads to training runs that are dominated by the overhead
of data ingest.  \footnote{Note that for the purposes of this test we could have
  shuffled the data and repacked the HDF5 files to provide a random distribution
  per file, but that optimization is infeasible in real scientific workflows.
  Additionally, it would not solve the problem of requiring a unique
  distribution for each subsequent epoch.}

\begin{figure}
  \centering
  \includegraphics[width=0.5\textwidth]{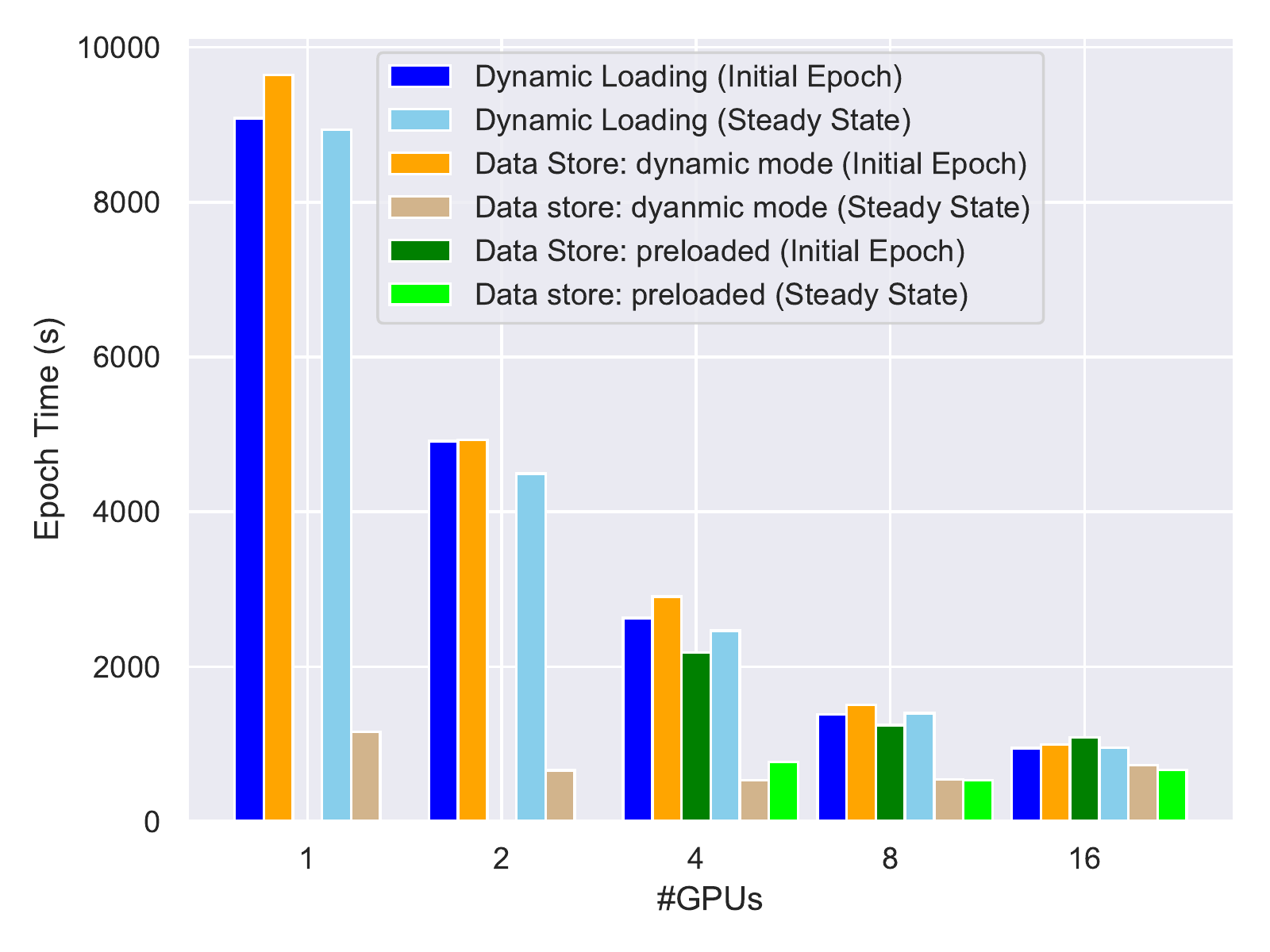}
  \caption{Evaluating training the CycleGAN model with the data store in dynamic
    mode, preloaded and without the data store on $1M$ sample data set.}
  \label{fig:data_store}
\end{figure}

\newcommand{\datastoreONE}{7.73}
\newcommand{\datastoreFOUR}{1.31}
\newcommand{\datastorePREfour}{1.43}
\newcommand{\datastoreVSpreFOUR}{1.10}

As described in Section~\ref{sec:lbann:datastore}, we have developed an
in-memory data store that caches the training, evaluation, and potentially test
data sets in the system's host memory.  Figure \ref{fig:data_store} show the
result of increasing the data parallelism by varying the number of nodes used by
the trainer using a $1M$ sample data set on the Lassen system.  On the x-axis is
the number of GPUs used to train the model and the y-axis is the epoch time in
seconds.  The results are presented for three configurations, each of which shows
the results for the inital and steady state epochs.  In Figure
\ref{fig:data_store} for each number of GPUs on the x-axis, the
left three bars show the initial time to load the
data set and complete the first epoch for each configuration.  The right three bars are the cost of each
subsequent epoch as the system enters steady state.  Note that the
configurations with the preloaded
data store did not have sufficient memory to load the model with 1 or 2 GPUs.
We see from Figure \ref{fig:data_store} that using the data store has
substantial benefit ranging from a massive ${\datastoreONE}\times$ for a trainer
using a single GPU to a ${\datastoreFOUR}\times$ for a trainer with 4 nodes.
When training with a fairly small data set like the 1M subset for this
experiment, preloading the data has limited advantage over the dynamically
loaded data store. Figure \ref{fig:data_store} shows that for 4 nodes there is a
${\datastorePREfour}\times$ improvement versus no data store, and a
${\datastoreVSpreFOUR}\times$ improvement over the dynamically loaded data
store.  However, the advantage of preloading substantially improved when transitioning to
the $10M$ sample data set due to the larger number of random file access required
by the dynamically loaded data store.  Therefore for the next experiment we
focus on using 4 nodes per trainer and preloaded data store.

\subsection{LTFB at scale}

\begin{figure}[ht]
  \centering
  {\includegraphics[width=0.49\textwidth]{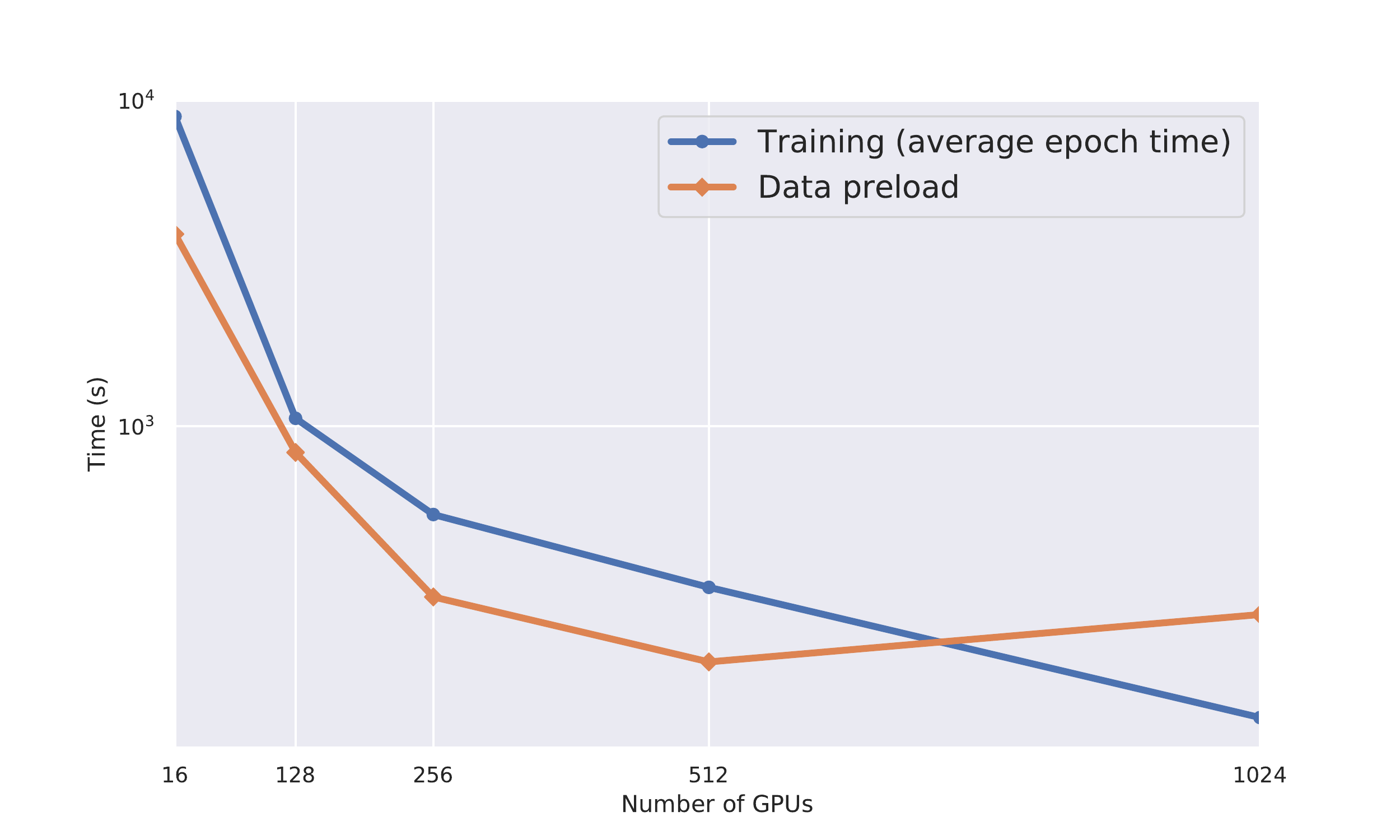}}
  \caption{LTFB training times for CycleGAN network on $10M$ sample data set}
  \label{fig:ltfb_strong_scaling}
\end{figure}

The previous two experiments demonstrate that data parallelism within a single
trainer can significantly improve the time required for data ingestion, but
performance will quickly plateau when using a fixed size mini-batch.
\footnote{As noted previously, large-scale mini-batch training is a regime that
  is compatible with the LTFB algorithm but does require significant tuning of
  the learning rate and has not been shown to generalize well to all problem
  domains.}  This experiment shows that LTFB is able to strong scale the
training time of a model without a loss of generalization.  We use a $10M$
sample sub-set of the entire $11M$ sample data set (leaving $1M$ for validation)
that is described in Section~\ref{sec:scienceml}, and scale up to 1024 Volta GPUs on
256 nodes of the Lassen system.

\newcommand{\LTFBscaling}{70.2}
\newcommand{\LTFBparallelEfficiency}{109}

In Figure \ref{fig:ltfb_strong_scaling} we show the per-epoch training time in
steady-state as we increase the number of LTFB trainers from 1 up to 64.  The x-axis is the number of GPUs used
across all trainers, with each trainer using 16 GPUs across 4 nodes.  This
corresponds to 1, 8, 16, 32, and 64 trainers in the experiment of Figure \ref{fig:ltfb_strong_scaling}  
Note that with 4 nodes per trainer, the data store of a single trainer was
unable to load the entire $10M$
sample training data set due to memory capacity.  Therefore for the single
trainer case we used 16 nodes per trainer and only 1 GPU per node.  This
increased memory capacity allowed the data store to load both the training and
validation data.  As noted previously, for the rest of the configurations, we
used 4 nodes and 16 GPUs per trainer.
As the number of trainers increase, the
steady-state epoch time decreases because each trainer is responsible for a
smaller partition of the data set.  Figure \ref{fig:ltfb_strong_scaling} shows
that 64 trainers achieve a speedup of ${\LTFBscaling}\times$ over the 1 trainer
baseline, and an effective ${\LTFBparallelEfficiency}\%$ parallel efficiency
with no loss in model quality.
Such superlinear speedups in strong scaling are due to ``cache'' effects as the
aggregate working set size is increased and demonstrates non-linear performance
improvements. Looking at the time required for preloading the data in Figure
\ref{fig:ltfb_strong_scaling} we see that at 64 trainers, the total time for all
trainers to load the data has degraded over the 32 trainer test point.  This
loss in performance is due to contention at the GFPS parallel
file system resulting from inter-trainer interference and will be addressed in future work.

\begin{figure}
  \centering
  \includegraphics[width=0.5\textwidth]{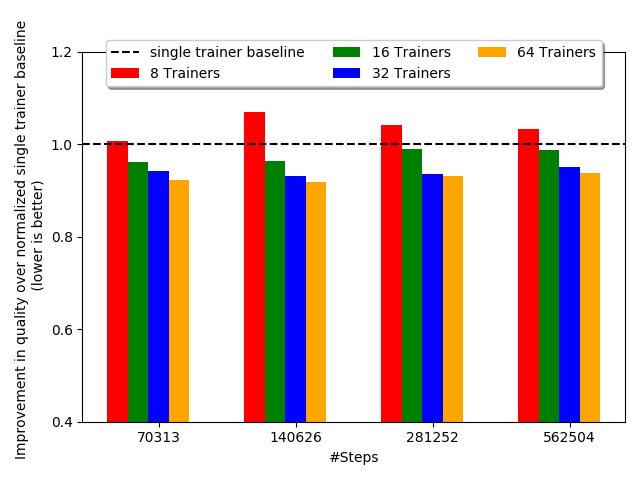}
  \caption{Improvement in quality (validation loss) over single-trainer baseline at different iterations (steps) per-trainer count.}
  \label{fig:quality_vs_steps}
\end{figure}

As noted earlier, the key attribute of the LTFB training algorithm is that the
quality of these models actually improves with the number of trainers.  While
each trainer is responsible for a smaller share of the global data set, the
tournament and model exchange in LTFB produces a model that has been trained on
a sufficient subset of the data to provide good generalizability (as measured
by forward and inverse loss on global validation data set). 
Figure \ref{fig:quality_vs_steps}  shows improvement in quality normalized over single-trainer baseline at the different
per-trainer iteration.  
Figure\ref{fig:quality_vs_steps} in combination with Figures \ref{fig:ltfb_strong_scaling}  
show that LTFB does not suffer quality
degradation with improved parallel performance. LTFB at bigger trainer sizes shows improved learning
quality and time to solution if measured by per-trainer number of iterations (steps) which roughly
translates to wall clock time.

Figures \ref{fig:gan_network_images} and \ref{fig:ltfb_pred_scalar} show ground truth
and predicted images and 15-D scalars of selected samples from the 1 million
validation dataset. These two figures show the quality of a trained multi-modal surrogate model
that jointly predict {\it realistic} images and scalars. The plots show
predicted images from different viewpoints and scalar values that closely
correlate with ground truth values.

\subsection{Comparing LTFB vs partitioned K-independent training}

\begin{figure*}
  \centering
  {\includegraphics[width=0.9\textwidth]{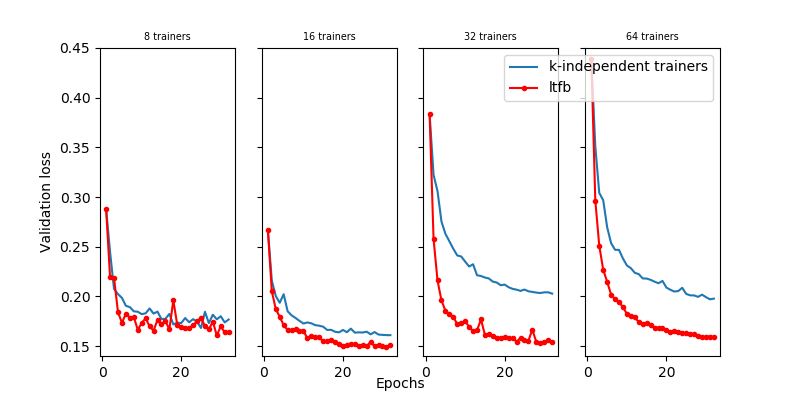}}
  \caption{Validation loss comparison of LTFB vs partitioned K-independent
training for CycleGAN (lower is better)}
  \label{fig:ltfb_vs_Kindependent}
\end{figure*}

One simpler alternative to LTFB would be to train K-independent models and simply select the best final result.
However, that assumes that a single trainer can hold the entire dataset during
training which is typically not the case.
For example, we were not able to process the data with only four trainers (using 4 nodes per trainer) as a quarter of the dataset proved too large.
As the number of trainers used increases, each trainer see fewer samples thus
making the data set size more manageable --
identical to the behavior of the LBANN trainers during the LTFB.
However, while the LTFB model exchange allows each trainer to create models that
have learned on compositions of the entire data set, every trainer in the
K-independent algorithm is restricted to an ever diminising portion of the
full data set.  As a result, the models created by the K-independent algorithm
generalize more poorly on a held out validation set.

To demonstrate this trade-off, \autoref{fig:ltfb_vs_Kindependent} shows comparisons
between running LTFB with $k$ trainers vs.\ $k$ independent trainers using a random $1/k$ subset of the data.
This compares roughly equal runtimes (i.e.\ equal number of iterations) and equal memory footprints between the two approaches.
As shown in the figure, the LTFB approach consistently achieves better results
in validation loss.
More importantly, with increasing $k$ the gap widens, which is unsurprising as the independent models see progressively smaller portions of the training data and thus generalize less well.
This means that for small data in which we can hold all or a significant portion
of the training data in memory the K-independent training approach is viable.
However, as the size of data sets increases the K-independent training approach breaks down.
Another viewpoint to consider is the likelihood of any one independent trainer
selecting a small subset of the training data that is distributed sufficiently well to lead to a generalizable model.
As the data becomes larger, our chances for this oracular selection will drastically decrease and in the
limit they may not even exist.  
Using the LTFB training algorithm instead produces models with better results
for the same compute resources and the datasize is only limited by the memory capacity of the entire system rather than that of a single trainer.
\section{Related Work}
\label{sec:related}
Recent successes of deep learning in many application domains have
encouraged renewed interest in understanding and optimizing neural
networks. Of particular interest and related to our work is the
subject of scaling neural network training on different architectural
platforms. Distributed and parallel training of deep learning
algorithms is well-studied but mostly focused on the traditional
fields of computer vision and speech recognition
~\cite{tensorflow2015-whitepaper},~\cite{strom2015scalable},~\cite{le2012building},~\cite{dean2012large}.
~\cite{recht2011hogwild,coates2013deep},
~\cite{iandola2015firecaffe},~\cite{goyal2017accurate,keskar2017large,cho2017}. Data and
model parallelism in these work are orthogonal to our work and are used within each
trainer to maximize parallelism. Here, we leverage ideas from these existing domains but focus on 
generative surrogate model for fusion science application and target HPC systems.

The LTFB algorithm was initially proposed in~\cite{Jacobs2017} and
demonstrates modest scaling on image classification benchmarks.  At
about the same time, researchers at Google Deepmind explored a similar idea for
hyperparameter exploration without data
partitioning~\cite{DBLP:journals/corr/abs-1711-09846}. Deepmind's
work did not address scaling but demonstrate the utility of population-based 
training approaches for hyperparameter optimization on a number of computer vision and
reinforcement learning benchmarks.  
In this work we present new methods for applying these tournmanet methods to
generative adversarial networks, extend them to a
different (scientific) domain with multivariate data, and provide a primary focus on large scale
training using HPC systems.

Scientific machine learning is drawing attention to problems that are beyond
computer vision and natural language processing \cite{spears-deeplearning,
DBLP:journals/corr/KarpatneAFSBGSS16}. 
Prominent recent works are: exascale deep learning for climate
analytics ~\cite{kurth2018exascale}, and $CosmoFlow$ that used deep learning to learn the universe
at scale ~\cite{mathuriya2018cosmoflow}.  In both
cases, straight forward large-batch data parallelism is used without
addressing convergence issues or the quality of trained regression
network

The climate analytics and cosmology problems mentioned above emphasize the
challenges of data ingestion and present
novel opportunities to improve performance. Kurth et al.~\cite{kurth2018exascale}
developed distributed data staging system in which each rank first
copies a disjoint subset of the entire dataset from the parallel file system
into its fast local storage, and then distributes the files to other ranks that
require them using the point-to-point MPI communication.
This significantly reduces the I/O bottleneck at the filesystem.
However, as the data staging precedes training, each rank needs to hold all the
data files it needs to access on the local storage.
On the other hand, LBANN offers a built-in, in-memory data store in which sample
data are transferred from the owner rank to the consumer rank as needed before
every minibatch using non-blocking communication.
This eliminates the redundant in-memory copies of data, hides the overhead in
redistributing them and reduces the volume of the redistribution
in case that each data file contains multiple samples.

\section{Conclusions}
\label{sec:concl}

In this paper we have presented novel improvements to the ``Let a Thousand Flowers
Bloom'' (LTFB) torunament algorithm that allow it to train traditional and
generative adversarial networks at scale.  We have discussed how it is
integrated into the LBANN open-source scalable deep learning framework, and
demonstrated how it enables scalable training of a massive scientific data set.
We have accelerated the training and discovery of novel neural network
architectures for simulations of capsule implosions for inertial confinement
fusion.  Our tournament algorithm allows for the parallel training of a single
neural network model that has learned on multiple silos of a partitioned data
set.  We show that as the number of parallel trainers increases, and thus the
percentage of the data set show to each trainer decreases, LTFB enables strong
scaling of the training time without loss of network quality.  We are working
with a very large scientific data set that has 10 million unique training
samples with 120 million images and 150 million scalar values.  To enable efficient
parallel training we have developed an in-memory distributed data store that
significantly improves the steady state per epoch training time.  The
aggregation of these technologies provides a scalable deep learning framework
that is extremely well suited for parallel training on massive data sets.  These
capabilities provide the first step along the path of developing a cognitive
simulation workflow.

\section*{Acknowledgment}

This work was performed under the auspices
   of the U.S. Department of Energy by Lawrence Livermore National
   Laboratory under Contract DE-AC52-07NA27344 (LLNL-CONF-677443).
   Funding partially provided by LDRD 18-SI-002.

\bibliographystyle{IEEEtran}
\bibliography{lbann}

\begin{thebibliography}{10}
\providecommand{\url}[1]{#1}
\csname url@samestyle\endcsname
\providecommand{\newblock}{\relax}
\providecommand{\bibinfo}[2]{#2}
\providecommand{\BIBentrySTDinterwordspacing}{\spaceskip=0pt\relax}
\providecommand{\BIBentryALTinterwordstretchfactor}{4}
\providecommand{\BIBentryALTinterwordspacing}{\spaceskip=\fontdimen2\font plus
\BIBentryALTinterwordstretchfactor\fontdimen3\font minus
  \fontdimen4\font\relax}
\providecommand{\BIBforeignlanguage}[2]{{%
\expandafter\ifx\csname l@#1\endcsname\relax
\typeout{** WARNING: IEEEtran.bst: No hyphenation pattern has been}%
\typeout{** loaded for the language `#1'. Using the pattern for}%
\typeout{** the default language instead.}%
\else
\language=\csname l@#1\endcsname
\fi
#2}}
\providecommand{\BIBdecl}{\relax}
\BIBdecl

\bibitem{spears-deeplearning}
\BIBentryALTinterwordspacing
B.~K. Spears, J.~Brase, P.-T. Bremer, B.~Chen, J.~Field, J.~Gaffney, M.~Kruse,
  S.~Langer, K.~Lewis, R.~Nora, J.~L. Peterson, J.~Jayaraman~Thiagarajan,
  B.~Van~Essen, and K.~Humbird, ``Deep learning: A guide for practitioners in
  the physical sciences,'' \emph{Physics of Plasmas}, vol.~25, no.~8, p.
  080901, 2018. [Online]. Available: \url{https://doi.org/10.1063/1.5020791}
\BIBentrySTDinterwordspacing

\bibitem{gaffney2014thermodynamic}
J.~Gaffney, P.~Springer, and G.~Collins, ``Thermodynamic modeling of
  uncertainties in {NIF ICF} implosions due to underlying microphysics
  models,'' in \emph{APS Meeting Abstracts}, 2014.

\bibitem{springer2013integrated}
P.~Springer, C.~Cerjan, R.~Betti, J.~Caggiano, M.~Edwards, J.~Frenje, V.~Y.
  Glebov, S.~Glenzer, S.~Glenn, N.~Izumi \emph{et~al.}, ``Integrated
  thermodynamic model for ignition target performance,'' in \emph{EPJ Web of
  Conferences}, vol.~59.\hskip 1em plus 0.5em minus 0.4em\relax EDP Sciences,
  2013, p. 04001.

\bibitem{van2015lbann}
\BIBentryALTinterwordspacing
B.~Van~Essen, H.~Kim, R.~Pearce, K.~Boakye, and B.~Chen, ``{LBANN}: {Livermore}
  big artificial neural network {HPC} toolkit,'' in \emph{Proceedings of the
  Workshop on Machine Learning in High-Performance Computing Environments
  (MLHPC)}, 2015. [Online]. Available:
  \url{https://doi.org/10.1145/2834892.2834897}
\BIBentrySTDinterwordspacing

\bibitem{lbann2019}
\BIBentryALTinterwordspacing
T.~Benson, N.~Dryden, R.~Forsyth, D.~Hysom, S.~A. Jacobs, N.~Maruyama,
  D.~McKinney, A.~Moody, T.~Moon, Y.~Oyama, J.-S. Yeom, A.~Yoo, and
  B.~Van~Essen, ``{L}ivermore {B}ig {A}rtificial {N}eural {N}etwork ({LBANN})
  {T}oolkit,'' 2019. [Online]. Available: \url{https://github.com/LLNL/lbann}
\BIBentrySTDinterwordspacing

\bibitem{hydrogen2019}
\BIBentryALTinterwordspacing
T.~Benson, T.~Moon, J.~Poulson, and B.~Van~Essen, ``{H}ydrogen {D}istributed
  {L}inear {A}lgebra {L}ibrary,'' 2019. [Online]. Available:
  \url{https://github.com/LLNL/Elemental}
\BIBentrySTDinterwordspacing

\bibitem{dryden2018}
N.~Dryden, N.~Maruyama, T.~Moon, T.~Benson, A.~Yoo, M.~Snir, and B.~Van~Essen,
  ``Aluminum: An asynchronous, {GPU}-aware communication library optimized for
  large-scale training of deep neural networks on {HPC} systems,'' in
  \emph{Proceedings of the Workshop on Machine Learning in HPC Environments
  (MLHPC)}, 2018.

\bibitem{aluminum2019}
\BIBentryALTinterwordspacing
N.~Dryden, N.~Maruyama, T.~Moon, T.~Benson, A.~Yoo, and B.~Van~Essen,
  ``Aluminum {GPU}-aware communication library,'' 2019. [Online]. Available:
  \url{https://github.com/LLNL/Aluminum}
\BIBentrySTDinterwordspacing

\bibitem{betti2015alpha}
R.~Betti, A.~Christopherson, B.~Spears, R.~Nora, A.~Bose, J.~Howard, K.~Woo,
  M.~Edwards, and J.~Sanz, ``Alpha heating and burning plasmas in inertial
  confinement fusion,'' \emph{Physical review letters}, vol. 114, no.~25, p.
  255003, 2015.

\bibitem{marinak01}
\BIBentryALTinterwordspacing
M.~M. Marinak, G.~D. Kerbel, N.~A. Gentile, O.~Jones, D.~Munro, S.~Pollaine,
  T.~R. Dittrich, and S.~W. Haan, ``Three-dimensional hydra simulations of
  national ignition facility targets,'' \emph{Physics of Plasmas}, vol.~8,
  no.~5, pp. 2275--2280, May 2001. [Online]. Available:
  \url{http://dx.doi.org/10.1063/1.1356740}
\BIBentrySTDinterwordspacing

\bibitem{doi:10.1063/1.4977912}
\BIBentryALTinterwordspacing
J.~L. Peterson, K.~D. Humbird, J.~E. Field, S.~T. Brandon, S.~H. Langer, R.~C.
  Nora, B.~K. Spears, and P.~T. Springer, ``Zonal flow generation in inertial
  confinement fusion implosions,'' \emph{Physics of Plasmas}, vol.~24, no.~3,
  p. 032702, 2017. [Online]. Available: \url{https://doi.org/10.1063/1.4977912}
\BIBentrySTDinterwordspacing

\bibitem{kailkhura2018spectral}
B.~Kailkhura, J.~J. Thiagarajan, C.~Rastogi, P.~K. Varshney, and P.-T. Bremer,
  ``A spectral approach for the design of experiments: Design, analysis and
  algorithms,'' \emph{The Journal of Machine Learning Research}, vol.~19,
  no.~1, pp. 1214--1259, 2018.

\bibitem{Goodfellow2014}
\BIBentryALTinterwordspacing
I.~Goodfellow, J.~Pouget-Abadie, M.~Mirza, B.~Xu, D.~Warde-Farley, S.~Ozair,
  A.~Courville, and Y.~Bengio, ``Generative adversarial nets,'' in
  \emph{Advances in Neural Information Processing Systems 27}, Z.~Ghahramani,
  M.~Welling, C.~Cortes, N.~D. Lawrence, and K.~Q. Weinberger, Eds.\hskip 1em
  plus 0.5em minus 0.4em\relax Curran Associates, Inc., 2014, pp. 2672--2680.
  [Online]. Available:
  \url{http://papers.nips.cc/paper/5423-generative-adversarial-nets.pdf}
\BIBentrySTDinterwordspacing

\bibitem{osti_1510714}
\BIBentryALTinterwordspacing
R.~Anirudh, P.-T. Bremer, J.~J. Thiagarajan, and {USDOE National Nuclear
  Security Administration}, ``Cycle consistent surrogate for inertial
  confinement fusion,'' 2 2019. [Online]. Available:
  \url{https://www.osti.gov//servlets/purl/1510714}
\BIBentrySTDinterwordspacing

\bibitem{Jacobs2017}
\BIBentryALTinterwordspacing
S.~A. Jacobs, N.~Dryden, R.~Pearce, and B.~Van~Essen, ``Towards scalable
  parallel training of deep neural networks,'' in \emph{Proceedings of the
  Machine Learning on HPC Environments}, ser. MLHPC'17.\hskip 1em plus 0.5em
  minus 0.4em\relax New York, NY, USA: ACM, 2017, pp. 5:1--5:9. [Online].
  Available: \url{http://doi.acm.org/10.1145/3146347.3146353}
\BIBentrySTDinterwordspacing

\bibitem{poulson2013elemental}
J.~Poulson, B.~Marker, R.~A. Van~de Geijn, J.~R. Hammond, and N.~A. Romero,
  ``Elemental: A new framework for distributed memory dense matrix
  computations,'' \emph{ACM Transactions on Mathematical Software}, vol.~39,
  no.~2, p.~13, 2013.

\bibitem{conduit2019}
\BIBentryALTinterwordspacing
C.~Harrison, B.~Whitlock, and J.~Ciurej, ``Conduit:{S}implified {D}ata
  {E}xchange for {HPC} {S}imulations,'' 2019. [Online]. Available:
  \url{https://github.com/LLNL/conduit}
\BIBentrySTDinterwordspacing

\bibitem{tensorflowDataCache}
\BIBentryALTinterwordspacing
``{T}ensor{F}low: {D}ata {I}nput {P}ipeline,'' 2019. [Online]. Available:
  \url{https://www.tensorflow.org/guide/performance/datasets}
\BIBentrySTDinterwordspacing

\bibitem{pytorchDataLoader}
\BIBentryALTinterwordspacing
``{P}y{T}orch: {D}ata{L}oader,'' 2019. [Online]. Available:
  \url{https://pytorch.org/docs/stable/data.html}
\BIBentrySTDinterwordspacing

\bibitem{DBLP:journals/corr/abs-1711-09846}
\BIBentryALTinterwordspacing
M.~Jaderberg, V.~Dalibard, S.~Osindero, W.~M. Czarnecki, J.~Donahue, A.~Razavi,
  O.~Vinyals, T.~Green, I.~Dunning, K.~Simonyan, C.~Fernando, and
  K.~Kavukcuoglu, ``Population based training of neural networks,''
  \emph{CoRR}, vol. abs/1711.09846, 2017. [Online]. Available:
  \url{http://arxiv.org/abs/1711.09846}
\BIBentrySTDinterwordspacing

\bibitem{DBLP:journals/corr/DurugkarGM16}
\BIBentryALTinterwordspacing
I.~P. Durugkar, I.~Gemp, and S.~Mahadevan, ``Generative multi-adversarial
  networks,'' \emph{CoRR}, vol. abs/1611.01673, 2016. [Online]. Available:
  \url{http://arxiv.org/abs/1611.01673}
\BIBentrySTDinterwordspacing

\bibitem{tensorflow2015-whitepaper}
\BIBentryALTinterwordspacing
M.~Abadi, A.~Agarwal, P.~Barham, E.~Brevdo, Z.~Chen, C.~Citro, G.~S. Corrado,
  A.~Davis, J.~Dean, M.~Devin, S.~Ghemawat, I.~Goodfellow, A.~Harp, G.~Irving,
  M.~Isard, Y.~Jia, R.~Jozefowicz, L.~Kaiser, M.~Kudlur, J.~Levenberg,
  D.~Man\'{e}, R.~Monga, S.~Moore, D.~Murray, C.~Olah, M.~Schuster, J.~Shlens,
  B.~Steiner, I.~Sutskever, K.~Talwar, P.~Tucker, V.~Vanhoucke, V.~Vasudevan,
  F.~Vi\'{e}gas, O.~Vinyals, P.~Warden, M.~Wattenberg, M.~Wicke, Y.~Yu, and
  X.~Zheng, ``{TensorFlow}: Large-scale machine learning on heterogeneous
  systems,'' 2015. [Online]. Available: \url{https://www.tensorflow.org/}
\BIBentrySTDinterwordspacing

\bibitem{strom2015scalable}
N.~Strom, ``Scalable distributed {DNN} training using commodity {GPU} cloud
  computing,'' in \emph{INTERSPEECH}, vol.~7, 2015, p.~10.

\bibitem{le2012building}
Q.~V. Le, R.~Monga, M.~Devin, K.~Chen, G.~S. Corrado, J.~Dean, and A.~Y. Ng,
  ``Building high-level features using large scale unsupervised learning,'' in
  \emph{In International Conference on Machine Learning, 2012. 103}, 2012.

\bibitem{dean2012large}
J.~Dean, G.~Corrado, R.~Monga, K.~Chen, M.~Devin, M.~Mao, A.~Senior, P.~Tucker,
  K.~Yang, Q.~V. Le \emph{et~al.}, ``Large scale distributed deep networks,''
  in \emph{Advances in Neural Information Processing Systems}, 2012.

\bibitem{recht2011hogwild}
B.~Recht, C.~Re, S.~Wright, and F.~Niu, ``Hogwild: A lock-free approach to
  parallelizing stochastic gradient descent,'' in \emph{Advances in Neural
  Information Processing Systems}, 2011, pp. 693--701.

\bibitem{coates2013deep}
A.~Coates, B.~Huval, T.~Wang, D.~Wu, B.~Catanzaro, and N.~Andrew, ``Deep
  learning with {COTS} {HPC} systems,'' in \emph{International Conference on
  Machine Learning (ICML)}, 2013.

\bibitem{iandola2015firecaffe}
F.~N. Iandola, K.~Ashraf, M.~W. Moskewicz, and K.~Keutzer, ``{FireCaffe}:
  Near-linear acceleration of deep neural network training on compute
  clusters,'' \emph{arXiv preprint arXiv:1511.00175}, 2015.

\bibitem{goyal2017accurate}
P.~Goyal, P.~Doll{\'a}r, R.~Girshick, P.~Noordhuis, L.~Wesolowski, A.~Kyrola,
  A.~Tulloch, Y.~Jia, and K.~He, ``Accurate, large minibatch {SGD}: training
  {ImageNet} in 1 hour,'' \emph{arXiv preprint arXiv:1706.02677}, 2017.

\bibitem{keskar2017large}
N.~S. Keskar, D.~Mudigere, J.~Nocedal, M.~Smelyanskiy, and P.~T.~P. Tang, ``On
  large-batch training for deep learning: Generalization gap and sharp
  minima,'' in \emph{Proceedings of the Fifth International Conference on
  Learning Representations}, 2017.

\bibitem{cho2017}
M.~{Cho}, U.~{Finkler}, S.~{Kumar}, D.~{Kung}, V.~{Saxena}, and D.~{Sreedhar},
  ``{PowerAI DDL},'' \emph{ArXiv e-prints}, Aug. 2017.

\bibitem{DBLP:journals/corr/KarpatneAFSBGSS16}
\BIBentryALTinterwordspacing
A.~Karpatne, G.~Atluri, J.~H. Faghmous, M.~Steinbach, A.~Banerjee, A.~R.
  Ganguly, S.~Shekhar, N.~F. Samatova, and V.~Kumar, ``Theory-guided data
  science: {A} new paradigm for scientific discovery,'' \emph{CoRR}, vol.
  abs/1612.08544, 2016. [Online]. Available:
  \url{http://arxiv.org/abs/1612.08544}
\BIBentrySTDinterwordspacing

\bibitem{kurth2018exascale}
T.~Kurth, S.~Treichler, J.~Romero, M.~Mudigonda, N.~Luehr, E.~Phillips,
  A.~Mahesh, M.~Matheson, J.~Deslippe, M.~Fatica \emph{et~al.}, ``Exascale deep
  learning for climate analytics,'' in \emph{Proceedings of the International
  Conference for High Performance Computing, Networking, Storage, and
  Analysis}, 2018.

\bibitem{mathuriya2018cosmoflow}
A.~Mathuriya, D.~Bard, P.~Mendygral, L.~Meadows, J.~Arnemann, L.~Shao, S.~He,
  T.~Karna, D.~Moise, S.~J. Pennycook \emph{et~al.}, ``{CosmoFlow}: Using deep
  learning to learn the universe at scale,'' in \emph{Proceedings of the
  International Conference for High Performance Computing, Networking, Storage,
  and Analysis}, 2018.

\end{thebibliography}

\end{document}